\documentclass[preprint,nofootinbib,showkeys]{revtex4-2}
\usepackage[T1]{fontenc}
\usepackage[latin9]{inputenc}
\usepackage{verbatim}
\usepackage{amsmath}
\usepackage{amstext}
\usepackage{color}
\usepackage{graphicx,psfrag}
\usepackage{amssymb}

\def\eg{\emph{e.g.~}}
\def\ie{\emph{i.e.~}}

\def\beq{\begin{equation}}
\def\eeq{\end{equation}}
\def\bea{\begin{eqnarray}}
\def\eea{\end{eqnarray}}
\def\beg{\begin{lyxgreyedout}}
\def\eeg{\end{lyxgreyedout}}
\def\nn{\nonumber}

\begin{document}

\title{New Stringy Physics beyond Quantum Mechanics \\ from the Feynman Path Integral }

\author{Donatello Dolce}

\affiliation{ University of Camerino, Piazza Cavour 19F, 62032 Camerino, Italy}

\begin{abstract}
%\abstract{
By investigating the Feynman Path Integral we prove that elementary quantum particle dynamics are directly associated to single compact  (cyclic) world-line parameters, playing the role of  the particles' internal clock, implicit in ordinary undulatory mechanics and indirectly observed for instance in Time Crystals. This allows us to formulate a novel purely four-dimensional stringy description of elementary particles as possible physics beyond quantum mechanics. The novelty of this approach is that quantum mechanics originates from a non-trivial compact nature of the minkowskian space-time.  Our result is a further evidence in support of Elementary Cycles Theory (ECT),  which in previous papers has been proven  to be consistent with known physics from theoretical particle physics to condensed matter. Here we provide additional conceptual arguments in support to this novel unified scenario of quantum and relativistic physics, potentially deterministic, and fully falsifiable having no fine-tunable parameters. The first evidences of such new physics characterized by ultra-fast cyclic time dynamics will be observed by probing quantum phenomena with experimental time accuracy of the order of $10^{-21}$\textit{sec}.  Considerations about the emergence of the arrow of time from the realm of pure, zero temperature, quantum physics governed by intrinsic time periodicity are also provided. Concerning Einstein's dilemma ``God does not play dice'' we conclude that, all in all, ``God'' would have no fun playing quantum dice.
\end{abstract}


%\PACS{{Feynman Path Integral} \and {Physics beyond Quantum Mechanics} \and { Time Crystals } \and { Space-Time Geometry } \and {Elementary Particles } \and {Concept of Time}}
\maketitle

%\begin{keyword}
%\keywords{Feynman Path Integral, Physics beyond Quantum Mechanics, Time Crystals, Space-Time Geometry, Elementary Particles, Concept of Time}
%\end{keyword}

\newpage

\tableofcontents{}

\newpage

\section*{Introduction}

The Feynman Path Integral (FPI) is one of the most important theoretical achievements of modern physics. It represents a formulation of Quantum Mechanics (QM)  equivalent to the axiomatic one, as proven by Feynman \cite{Feynman:1942us}. Its  validity has been confirmed with impressing accuracy in countless experiments. In this paper we propose an analysis in which the FPI reveals an unedited,  possible, intrinsically compact nature of relativistic space-time beyond QM, \cite{Dolce:2009ce,Dolce:tune,Dolce:AdSCFT,Dolce:cycles,Dolce:2016rfc,Dolce:SuperC,Dolce:EPJP,Dolce:FQXi,Dolce:IntroEC,Dolce:ICHEP2012,Dolce:QTRF5,Dolce:TM2012}.  Nevertheless it must be clear that the correctness of the ordinary mathematical formulations of quantum physics in predicting probability amplitudes as well as the essence of relativity are definitively not questioned in this paper. 

Our analysis of the FPI seems to imply a physical principle at the origin of QM:  elementary particles are characterized by compact world-line dynamics at Compton scales. These result in the space-time recurrences of ordinary undulatory mechanics if observed in generic reference frames. We summarize our \textit{ansatz} by saying that \textit{elementary particles are the elementary clocks of nature}. This critical conceptual point has a renewed theoretical and phenomenological support in the recent discovery of Time Crystals \cite{Wilczek:2012jt,Sacha:2017}. The ordinary mathematical formulations of relativistic QM such as the FPI formulation investigated here or Quantum Field Theory (QFT) can be interpreted as affective descriptions. QM  emerges as  statistical, low time resolution, effective description of ultra-fast cyclic relativistic space-time dynamics at the base of wave-particle duality they can not be directly detected with the present experimental temporal resolution. 

In the standard interpretation of the FPI the probability amplitude of a quantum particle to travel between two space-time points is given by the interference of all the paths --- classical and non-classical --- joining them.  The probabilistic weight is given by the particle action  $ \mathcal  S$ evaluated along each path. 
Only one classical path is possible between two  end-points as long as the classical action is defined on the ordinary, implicitly non-compact, relativistic space-time.  
As pointed out by Feynman, the price to pay to reconcile QM with the Lagrangian formulation of classical mechanics --- defined in the ordinary non compact space-time --- is to give physical meaning to the non-classical paths, contrarily to the prescription of the classical least action principle. 

Nevertheless Feynman himself,  with his checkerboard model \cite{Feynman:100771}, tried to go beyond this interpretation by investigating the possibility to write the FPI as a discrete sum of paths, \ie paths labeled by integer numbers. 
If this is the case it would be possible to conceive a classical action whose minimization yields a numerable infinity of (degenerate) classical paths. For instance, in  a compact (or cyclic) geometry,  infinite degenerate classical paths are possible between two arbitrary end-points, due to the Boundary Conditions (BCs). These degenerate paths  would be labeled by an integer number (winding number). 

In par.(\ref{FPI:proof}) we will prove, based on widely accepted facts about QFT and QM, that the FPI can be equivalently written in terms of cyclic paths labelled by a winding number.  In par.({\ref{worldcycles}}) we will analyze the resulting compact (cyclic) word-line structure of elementary particles and its relationship  with the Compton periodicity at the base of de-Broglie wave-particle duality. This will naturally imply a stringy description of elementary particles as defined in par.(\ref{cyclicstrings}), named here Elementary Cycle String Theory (ECST). These novel strings constitute the basic quantum oscillators of ordinary quantum fields. Their target space-time turns out to be the ordinary four-dimensional Minkowskian space-time (without the problematic extradimensions of ordinary string theory) provided contravariant BCs which will encode the undulatory nature of elementary particles directly into the fabric of space-time, par.(\ref{space-time-cycles}). In order to show how relativistic causality is naturally satisfied by the theory, in par.(\ref{interactions}) we will investigate interactions, finding a common geometrodynamical description of both gauge and gravitational interactions. We will give predictions about new possible physics beyond QM,  par.(\ref{new:phys:BQM}), describing how ordinary QM can possibly emerge as statistical, low-time resolution, effective description of the inner ultra-fast cyclic dynamics associated to elementary particles, par.(\ref{emerging:QM}). Finally, by considering thermodynamical systems, we will shortly discuss the new elements allowed by the theory for a modern description of the entropic arrow of time, par.(\ref{arrow:time}).  

\section{Feynman Path Integral Analysis}\label{FPI:proof}

We now show that  the ordinary FPI of QM can be interpreted as an integral over a discrete sum of paths --- labelled by an integer number.  
Let us consider the FPI for a free relativistic scalar particle of momentum $\vec k$ traveling between two space-time end points ${x_i}^\mu$ and  ${x_f}^\mu$. In performing our analysis we will focus on the quantum field component $ \Phi_{\vec k}(x)$ of momentum $\vec k$. The ordinary quantum field $ \Phi(x)$ associated to the particle is obtained by integrating $ \Phi_{\vec k}(x)$ over momentum space. In other words we want to focalize our investigation on the fundamental quantum oscillators $ \Phi_{\vec k}(x)$ at the base of the quantum fields $ \Phi(x)$.  

 By denoting   the initial and final time of the evolution $t_i$ to $t_f$ respectively, the ordinary FPI  is defined as (we will exclusively use natural units $\hbar = c = 1$)
\beq\label{Feyn:Path:Int}
\mathcal Z = \int \mathcal D^3 x e^{i \mathcal  S^{} [t_f; t_i] }, 
\eeq
where $\mathcal S^{Free} [t_f; t_i] =\mathcal \int_{t_i}^{t_f}\mathcal L^{} dt $ is the classical-relativistic action and $\mathcal L^{} = \vec{\mathcal P} \cdot  \vec{ \dot x} - \mathcal H$ is the free particle classical-relativistic Lagrangian written in terms of the Hamiltonian  and momentum  operators, $ \mathcal H$ and  $\vec{\mathcal P} $, respectively. 

We perform the ordinary time slicing. In the limit  $N \rightarrow \infty$  we have 
\beq\label{Fey:Int:Slice}
\mathcal Z =  \lim_{N \rightarrow \infty} \int  \left ( \prod_{m=0}^{N-1}  d^3 x_m \right ) \mathcal  U_{f, N-1} %\mathcal  U_{N-1, N-2}  
\cdots  
  \mathcal U_{i,1}\,,
\eeq
where the elementary Feynman space-time evolutions are 
\bea\label{Fey:element}
\mathcal U_{m+1, m} &= & %\mathcal  U(\vec x_{m+1}, t_{m+1}; \vec x_m, t_m) = \langle \Phi_{\vec k}| e^{i \mathcal L^{Free} d t } | \Phi_{\vec k} \rangle  \nn \\ &=&  
\langle \Phi_{\vec k}| e^{-i [\mathcal H (\vec k) \Delta t_m - \vec {\mathcal P} \cdot  \Delta \vec x_m]} | \Phi_{\vec k} \rangle_H \,.
\eea 
The spatial intervals $\Delta \vec x_m =  \vec x_{m+1} - \vec x_{m}$ refer to the time slice $ \Delta t_m = t_{m+1} - t_m$. As  appearing in many textbook in introducing the FPI, $| \Phi_{\vec k} \rangle_H $ is the Heisenberg multiparticle quantum state\footnote{We address it as \textit{quantum particle} of momentum $\vec k$. It is the result of the field component $\Phi_{\vec k}$ second quantization and it is used. Not be confused with the single particle of ordinary QM.} of momentum $\vec k$ associated to the particle, defined in terms of the local energy eigenstates  $| n_{\vec k} \rangle$. 
Here we are including both positive and negatives frequencies $k_0 = \pm \omega(\vec k)/c$, according to the quadratic relativistic dispersion relation  $ M^2 =  k_0^2 -  \vec k^2 $ and ordinary QFT.

  It is well known that the second quantization of each field component $ \Phi_{\vec k}(x)$ %$| \Phi_{\vec k} \rangle$ 
  yields the normal ordered energy spectrum $\omega_{\tilde n_{\vec k}}(\vec k) :=: \tilde n_{\vec k} \omega(\vec k)$ with $\tilde n_{\vec k} \in \mathbb N$ for both the positive and negative frequencies.  As usual in FPI analysis %, and without loss of generality,  
we will always assume normal ordering %denoted by the colon symbol in the equivalence
 (represented by $:=:$) without affecting the general validity of our results.\footnote{The vacuum energy is retrieved by simply replacing everywhere the condition of periodicity (PBCs) with the condition of anti-periodicity, equally allowed by the variational principle (combination of Dirichlet and Neumann BCs) so that $\omega_{\tilde n_{\vec k}}(\vec k) = (\tilde n_{\vec k} + \frac{1}{2}) \omega(\vec k)$ with $\tilde n_{\vec k} \in \mathbb N$. All our results will be  equally valid. }
 So, when both positive and negative frequencies of $| \Phi_{\vec k} \rangle$ are considered, the energy spectrum of the Hamiltonian operator $\mathcal H (\vec k)$ in  eq.(\ref{Fey:element}) can be equivalently written as 
$ \mathcal H (\vec k) |n_{\vec k}\rangle  
:= :n_{\vec k} \omega(\vec k) |n_{\vec k}\rangle$, with $n_{\vec k} \in \mathbb Z$. % and  $| \Phi_{\vec k} \rangle = \sum_{n_{\vec k} } | n_{\vec k} \rangle$. 
Through the quadratic relativistic  dispersion relation the energy spectrum fixes the momentum spectrum of $| \Phi_{\vec k} \rangle $ as well. It follows that
$\vec {\mathcal P} |n_{\vec k}\rangle :=: n_{\vec k} \vec k |n_{\vec k}\rangle$,  with $n_{\vec k} \in \mathbb Z$.\footnote{ 
Notice that the negative values of $n_{\vec k} \in \mathbb Z$ provide a correct description of the negative negative energy eigenmodes and their momenta. For instance, in ordinary QFT the annihilation operator is defined in terms of the creation operator as $\mathbf a^\dagger(\vec k) = \mathbf a(- \vec k) $.}
 %For instance consider photons whose massless dispersion relation is $\omega(\vec k) = |\vec k|$, so that the normally ordered energy spectrum $\omega_{n_{\vec k}} = n_{\vec k} \omega$  implies the  momentum spectrum $\vec k_{n_{\vec k}} = n_{\vec k} \vec k$. 
 
%So far we have essentially recalled widely accepted identities of second quantization and theirs direct consequences. 
We evaluate eq.(\ref{Fey:element}) by using the energy and momentum eigenvalues described above for the Hamiltonian and momentum operators acting on $| \Phi_{\vec k} \rangle$. Then we use the Poisson summation $\sum_{n \in \mathbb Z} e^{-i n y} = 2 \pi \sum_{n'  \in \mathbb Z}\delta(y + 2 \pi n')$,  where $n, n' \in \mathbb Z$.  We find that the elementary Feynman evolutions eq.(\ref{Fey:element}) can be written as a sums of Dirac deltas, \textit{a.k.a.} Dirac comb,    
\bea\label{Fey:elem:sum}
\mathcal  U_{m+1, m} &=&  
  \sum_{n_m \in \mathbb Z} e^{-i n_m [\omega(\vec k) \Delta t_m - \vec k \cdot  \Delta \vec{x}_m]}  \nonumber \\
 &=&   2 \pi   \sum_{n'_m \in \mathbb Z}  \delta  \left(\omega(\vec k) \Delta t_m - \vec k \cdot \Delta \vec x_m + 2 \pi {n'_m}\right ).%\nonumber \\
\eea
%These Dirac deltas describes elementary space-time paths associated to the free particle evolution.  

We plug this result in the FPI, eq.(\ref{Fey:Int:Slice}), and we apply the Dirac delta property $\int d^3 x_m \delta (\vec x_{m+1} - \vec x_m) \delta (\vec x_m - \vec x_{m-1}) = \delta  (\vec x_{m+1} - \vec x_{m-1})$. We finally find that the ordinary FPI of a free relativistic quantum particle of momentum $\vec k$ can be explicitly expressed as the sum of space-time paths, represented by Dirac deltas, and labeled by the integer number $n'$:
%\begin{widetext}
\bea\label{Fey:Path:Sum}
\mathcal Z  &=& \lim_{N \rightarrow \infty} (2 \pi)^{N} \sum_{n'_0,  \dots, n'_{N-1} \in \mathbb Z}   \delta \left(\omega(\vec k)  (t_f - t_i) 	-  \vec k \cdot  (\vec x_f - \vec x_i)  + 2 \pi ( {n'_0}  + \dots +  {n'_{N-1}}) \right )  \nonumber \\
 &\equiv&   2 \pi   \sum_{n' \in \mathbb Z}   \delta   \left(\omega(\vec k) (t_f - t_i) - \vec k \cdot   (\vec x_f - \vec x_i) + 2 \pi {n'} \right ). %\nonumber \\
  \eea
By writing this result in  covariant notation  we finally get 
\beq\label{path:sum:spacetime}
\mathcal Z =  \int \mathcal D^3 x e^{i \mathcal  S}  \equiv  2 \pi \!\!   \sum_{n' \in \mathbb Z} \delta \Big ( \omega_\mu (x_f^\mu - x_i^\mu) + 2 \pi  {n'}  \Big )\,,  
\eeq
where  $ \omega_\mu$ is the particle  four-momentum (which for the moment being does not depend on the space-time point $x$, \ie it is \textit{global}, as we are in the free case);  $\Delta x^\mu = x_f^\mu - x_i^\mu$ is the interval between the final and the initial space-time points of the free particle evolution. 

 For an observer in the particle reference frame, independently on the field component of momentum $\vec k$ considered so far for the FPI, the previous relation can be expressed in terms of relativistic invariants 
\bea\label{FPI:paths:rest:intro}
\mathcal Z  = \int \mathcal D^3 x e^{i \mathcal  S^{}} &\equiv&   2 \pi  \sum_{n' \in \mathbb Z} \delta \left( M ({\tau_f} - {\tau_i}) + 2 \pi {n'} \right ) \nn \\
&\equiv&   T_C \sum_{n' \in \mathbb Z} \delta \left({\tau_f} - {\tau_i} + {n'} T_C \right ) \,, 
\eea
 where the particle Compton time $T_C = 2 \pi / M$ is defined by the particle mass $M$.  The initial and final word-line points of the particle quantum evolution are $\tau_i$ and $\tau_f$, respectively.

Notice that this demonstration for a free multiparticle state of momentum $\vec k$,  is based on widely accepted mathematical identities of second quantization and their direct consequences. %, in particular the energy and momentum spectra of a Klein-Gordon field component of momentum $\vec k$, 
By construction it can be generalized to interacting particles and fields, see par.(\ref{interactions}), as well as to the functional formulation of the FPI, as proven in \cite{Dolce:2009ce,Dolce:tune,Dolce:AdSCFT,Dolce:cycles,Dolce:2016rfc,Dolce:SuperC,Dolce:EPJP,Dolce:FQXi}, see also  \cite{Dolce:IntroEC,Dolce:ICHEP2012,Dolce:QTRF5,Dolce:TM2012}.

\section{Elementary World-Cycles}\label{worldcycles}

From a physical point of view the FPI clearly reveals in eq.(\ref{FPI:paths:rest:intro}) or eq.(\ref{path:sum:spacetime}) something very interesting about the nature of space-time beyond QM. The integer index $n'$ labeling the paths is manifestly a \textit{winding number}.   In particular,  in eq.(\ref{FPI:paths:rest:intro})  the Dirac deltas describe all the possible classical degenerate paths between arbitrary end-points  $\tau_i$ and $\tau_f$ on a \textit{compact} world-line $\tau$ of compactification length $T_C$ and Periodic BCs (PBCs). We will name it  elementary particles' \textit{world-cycle} or \textit{proper-time-cycle}, being a constraint of intrinsic periodicity in the proper-time  fixed by the particle mass according to the Compton relation.  With intrinsic periodicity we mean that the end points $\tau_i$ and $\tau_f$ are defined modulo factors $T_C$, see eq.(\ref{FPI:paths:rest:intro}).  Notice,  for instance,  that intrinsic time periodicity for ground state quantum systems has been recently observed in Time Crystals \cite{Wilczek:2012jt,Sacha:2017}. This novel phenomenon of condensed matter appears to be here generalized to elementary particle physics.   %As we will se this compact world-line structure of space-time is at the origin of ordinary undulatory dynamics of elementary particles. 
% --- besides Carbon Nanotubes and Superconductivity as noted in \cite{Dolce:SuperC,Dolce:EPJP}.

The subtle but essential novelty with respect to the ordinary undulatory description of QM  is that the wave-particle duality here originates from the intrinsically compact nature of the elementary particle world-lines, that is directly from the fabric of relativistic space-time.  
The parametrization of the particle evolution in terms of the \textit{world-cycle} (rather than on a implicitly \textit{non-compact} world-line) is,  after all, the real essence of the wave-particle duality and undulatory mechanics, see also par.(\ref{space-time-cycles}). We will see that the \textit{world-cycle} recurrence implies the ordinary space-time recurrences of QM.   

As stated by de Broglie throughout all his seminal PhD thesis \cite{Broglie:1924} at the origin of modern undulatory formulation of QM, ``\textit{to each elementary particle with proper mass $M$, one may associate a periodic phenomenon of Compton periodicity}'', or, in Penrose words \cite{Penrose:cycles}, ``\textit{any stable massive particle behaves as a very precise quantum clock, which ticks away with Compton periodicity}''. Moreover, we must consider that, according to Einstein \cite{Einstein:1923}, ``\textit{a clock is a periodic phenomenon so that what it happens in a period is identical to what happens in any other period}''. 

\textit{Elementary particles are thus the elementary clocks of nature}, as far as QM is concerned. 
There is nothing wrong in describing  quantum particles as intrinsic clocks ticking at Compton rates. It is implicitly done every time we write a wave function or a field in terms of phasors, as in QM. 
On the other hand, massless particles such as photons or gravitons  have infinite world-line compactification lengths  $T_{C_\gamma} \rightarrow \infty$. They can be imagined as ``\textit{frozen clocks}'' \cite{Penrose:cycles}. That is, massless particles have infinite world-lines as usual. So, they can be used to define  the ordinary \textit{reference} non-compact world-line(s) necessary for an observer to describe the relativistic motion of massive particles which, on the other hand, have compact \textit{world-cycles}.  This guarantees that the ordinary causal structure of relativistic physics will be preserved in the resulting description and massive particles can propagate in the whole space-time (defined by massless particles) despite their compact \textit{world-cycles}. A massive particle can be located at every generic point $\tau$ of the infinite world-line defined by massless particles, but its evolution  in that specific world-line point is characterized by a proper-time recurrence $T_C = 2 \pi / M$.  We will discuss these aspects in more detail in par.(\ref{interactions}).  For the moment being we can simply bear in mind that: $(a)$ it is a fact that the ultimate building blocks of the universe are elementary particles; $(b)$ it is a fact, as implicit in the wave-particle duality, that every elementary particle is a ``periodic phenomenon'' ticking at Compton rate \cite{Broglie:1924,Einstein:1923,Penrose:cycles}; $(a)$ $and$ $(b) \Rightarrow (c)$ it must be true that the whole physics can be consistently formulated in terms of elementary cycles --- and their modulations if interaction is concerned, as we will check. 

Naively, the fact that each elementary free particle is a persistent ``periodic phenomenon'' does not mean that all natural phenomena should be periodic in time, exactly as much as Newton's first law does not imply that everything in the universe should move in straight lines. The central point necessary to retrieve causality is that free particles imply \textit{persistent} periodicity whereas interacting particles imply \textit{locally} modulated periodicities (eg, \textit{eikonal}), see below. In the analogy with music also discussed in previous papers, pretending that intrinsic periodicity cannot describe the complexity of ordinary physics is like pretending that music instruments, being based on standing waves which are intrinsic periodic phenomena (PBCs in time), cannot reproduce elaborated symphonies. As for Newton's law our strategy is to clearly define the behavior of the isolated building-blocks of nature, and then generalize the description to interactions, par.(\ref{interactions}). It is sufficient to consider a simple system of two non-coherent periodic phenomena (rough description of a composite particle) to realize that intrinsic periodicity of elementary particles does not mean that everything in the universe should be periodic. In fact we get an ergodic system. When interactions and the consequent \textit{local} modulation of periodicities will be consider we will obtain the whole complexity of ordinary physical systems.

\section{Novel Stringy Description of Elementary Particles}\label{cyclicstrings}

Among the various interpretations of new physics beyond QM followed by eq.(\ref{FPI:paths:rest:intro}), in this paper we will focus on basic stringy aspects. They will constitute  \textit{Elementary Cycle String Theory} (ECST), which in turn is essentially based on Elementary Cycles Theory (ECT).  Even though the present paper is self-consistent, a wider and more formal view of the following description can be found in  \cite{Dolce:2009ce,Dolce:tune,Dolce:AdSCFT,Dolce:cycles,Dolce:2016rfc,Dolce:SuperC,Dolce:EPJP,Dolce:FQXi,Dolce:IntroEC,Dolce:ICHEP2012,Dolce:QTRF5,Dolce:TM2012}.  %The theory resulting from the condition of intrinsic periodicity of elementary particle that we will infer here directly from the FPI is essentially Elementary Cycles Theory (ECT).  

We have proven that the Feynman paths can be  interpreted as sum of degenerate classical  paths on a cyclic geometry. Now we will see that this has an equivalent description in terms of classical vibrating strings of novel type. 
It is easy to realize that the Feynman paths, written as in eq.(\ref{FPI:paths:rest:intro}),  actually are the degenerate classical solutions, from $\tau_i$ to $\tau_f$, making stationary the  action 
\beq\label{Action:free:cyclic:intro}
\mathcal S_{Compact} \dot = - \frac{1}{T_C} \oint^{\tau_f}_{(T_C), \tau_i} d \tau\,, %~~~(\text{compactification length $\lambda_C$ on $s$}),
\eeq
 where $\oint^{}_{(T_C)}$ means that  $\mathcal S_{Compact}$  is defined on the \textit{world-cycle} of Compton time periodicity $T_C = 2 \pi / M$. Not to be confused with $\mathcal  S$ appearing in FPI which is the ordinary classical-relativistic action and it is defined on the ordinary non-compact world-line. The end points $\tau_i$ and $\tau_f$ are therefore defined modulo factors $T_C$.  $\mathcal S_{Compact}$ will be regarded as the fundamental action of new physics beyond QM. Its classical solutions are all the Feynman paths of eq.(\ref{FPI:paths:rest:intro}).  $\mathcal  S$ is  the ordinary classical-relativistic action which however will be interpreted as the effective action emerging from $\mathcal S_{Compact}$ in the classical (no-quantum) limit.

We can easily see that $\mathcal S_{Compact}$ defined in eq.(\ref{Action:free:cyclic:intro}) is manifestly a \textit{string action}. It associates a closed string of novel type to each elementary particle of the Standard Model (SM) --- to each generic elementary particle of mass $M_q$ is associated a specific intrinsic world-line periodicity $T_{C_q} = 2 \pi / M_q$, whose duration depends on the particle mass. These novel relativistic strings, classical in the essence, constitute the fundamental quantum oscillators at the base of quantum fields,  as it can be easily seen from the demonstration of eq.(\ref{FPI:paths:rest:intro}).
%We name \textit{Elementary Cycle String Theory} (ECST) such stringy interpretation of new physics beyond QM inferred here from the FPI, eq.(\ref{FPI:paths:rest:intro}). ECST is explicitly based on ECT and  indirectly supported by the recent developments in Time Crystals.  

  The idea of a stringy description of elementary particles of course is not new, but surprisingly ECST, emerging from the FPI, is defined on a compact world-line (\textit{world-cycle}) rather than on the two dimensional world-sheet of \textit{Ordinary String Theory} (OST).  
From a historical point of view we know from Regge \textit{et al}. that the good mathematical properties of OST originates from the compact parameter of the ordinary two-dimensional world-sheet. Actually,  ECST inherits most of the mathematical beauty of OST\footnote{\textit{E.g.}, the vibrational harmonics coincides with the particle quantum excitations (Regge poles), the Fourier coefficients automatically satisfy Virasoro algebra and identify the Ladder operators of the ordinary Klein-Gordon field with implicit commutation relations (non second quantization is necessary) \cite{Dolce:AdSCFT}.}, as consequence of the compact world-line parameter \cite{Dolce:AdSCFT}. The non compact world-sheet parameter of OST was originally added to describe time evolution in the theory, before quantization. 
%As we have seen, there is nothing wrong in describing quantum particles in a \textit{world-cycles}. 
Our argument suggests that  the non-compact world-sheet parameter of OST is unnecessary to describe time evolution as soon as we take into account that elementary particles, the basic constituents of our universe, are the elementary clocks of nature according to QM. Particle physics and, in particular, particle's time evolutions, can be consistently formulated in terms of elementary cycles and their modulations.   The possibility of such a ECST is clear also if we take into account the novel developments in Time Crystals where the evolution of quantum systems, even if they are in their ground states, is characterized  by intrinsic periodicity in time, \ie BCs in the ordinary proper-time coordinate.

On the other hand, the single compact world-line parameter of ECST, contrarily to the ordinary two dimensional world-sheet, avoids problematic aspects of OST such as the proliferation of extra-dimensions.
As we are going to see, in ECST the target space-time is in fact the ordinary four-dimensional minkowskian space-time. The contravariant compactification at Compton lengths will encode undulatory relativistic mechanics directly into the structure of the four-dimensional minkowskian (flat) geometry of relativity.   Last but not least ECST, thanks to its four-dimensional target space-time, is phenomenological predictive. Actually, ECT has been successfully applied to describe quantum phenomena in different fields, from theoretical particle physics to condensed matter  \cite{Dolce:2009ce,Dolce:tune,Dolce:AdSCFT,Dolce:cycles,Dolce:2016rfc,Dolce:SuperC,Dolce:EPJP,Dolce:FQXi}. For instance, from the demonstration of par.(\ref{FPI:proof}) it is easy to see the link between such elementary strings and quantum fields.  The \textit{classical} elementary cycle strings can imagined as the fundamental \textit{quantum} oscillator of ordinary QFT, as it can be seen in the correspondence with the FPI derived above --- or in the derivation of the  axioms of QM  proven in \cite{Dolce:2009ce,Dolce:tune,Dolce:AdSCFT,Dolce:cycles,Dolce:2016rfc}.   

Here, for simplicity's sake, we have only considered  PBCs but,  in general, see \cite{Dolce:2009ce,Dolce:tune,Dolce:AdSCFT,Dolce:cycles,Dolce:2016rfc,Dolce:SuperC,Dolce:EPJP,Dolce:FQXi,Dolce:IntroEC,Dolce:ICHEP2012,Dolce:QTRF5,Dolce:TM2012}, all the combinations of Neumann and Dirichlet BCs are equally allowed by the variational principle for a bosonic relativistic theory (without additional boundary terms,  \cite{Csaki:2005vy,Henneaux:1992ig}). In fact, it is well-known in physics of the extra-dimensions, see \textit{e.g.} \cite{Csaki:2005vy,Henneaux:1992ig}, that relativistic dynamics are not broken by BCs as long as the BCs are allowed by  the variational principle for that specific relativistic action. 
For the fermionic relativistic action we have less trivial BCs resulting in ``twisted'' periodic dynamics. As discussed for instance  \cite{Dolce:cycles} the resulting generalization to fermions of our description has much in common with the \emph{zitterbewegung} \cite{Hestenes:zbw:1990,Wetterich_2021}. 
In general we can say that different types of particles (spin) correspond to different compactified geometries at Compton lengths.

\section{Elementary Space-Time Cycles}\label{space-time-cycles}

We now  consider the quantum evolution of a free particle of momentum $\vec k$  in a generic inertial reference frame. In this case the FPI can be decomposed in terms of a discrete sum of compact (cyclic) space-time  paths. In fact, according to the demonstration given in par.(\ref{FPI:proof}),  we have the equivalence eq.(\ref{path:sum:spacetime}) in covariant notation, 
where  $ \omega_\mu$ is the particle \textit{global} four-momentum. With the terms \textit{global} or \textit{persistent} we mean quantities that do not depend on the space-time point $x$ of the particle evolution, as the particle four-momentum of the free case. % In eq.(\ref{path:sum:spacetime}), $\Delta x^\mu = x_f^\mu - x_i^\mu$ is the interval between the final and the initial space-time points of the free particle evolution.  
% In the case of an inertial reference frame, which in our description necessarily refers to the case of a free relativistic quantum particle (the non-intertial one corresponds to interactions, see below), 

 In complete analogy with the rest frame description, eq.(\ref{path:sum:spacetime}) describes \textit{global} space-time cycles of  \textit{persistent} temporal recurrence $T = 2 \pi / \omega$ and \textit{persistent} spatial recurrences $\lambda^i = 2 \pi / k_i$, $i= 1, 2, 3$.  
 % Since we are in the free case the recurrences are global (persistent), that the particle can be located in every space-time point, but in that point the particle dynamics are characterized by tin sense that their recurrences are the same in every point in which the free particle field is defined, in the interaction case  they must be promoted to \textit{local} quantities, as we will see below. 
 %These are, of course, the ordinary recurrences of  QM --- imposed as constraints. That is, a particle can be located in every space-time point of the universe
 % but now they are ``super-imposed'' to the minkowskian space-time. 
In the free case they form the temporal and spatial components of a \textit{global} contravariant four-vector $\lambda^\mu = \{T,   \vec \lambda \}$. It is  fixed  by the  \textit{persistent} four-momentum $\omega_\mu =  \{\omega, - \vec k\}$ of the free  particle through the Planck constant, according to $ \omega_\mu \lambda^\mu = M T_C = 2 \pi$ (the fundamental topology of these space-time recurrences is $\mathbb S^1$). These are, of course, the ordinary  temporal and spatial recurrences of  QM --- imposed as space-time constraints --- as it is easily verified by Lorentz transforming the world-line Compton periodicity $T_C = 2 \pi / M $ to a generic inertial reference frame.  We have: $T_C = \gamma T - \gamma \vec \beta \cdot \vec \lambda$ where $\gamma = 1 / \sqrt{1 - \vec \beta^2} $ is the Lorentz factor, so that  $M T_C = \gamma M T - \gamma M \vec   \beta \cdot \vec  \lambda = \omega T - \vec k \cdot \vec \lambda = \omega_\mu \lambda^\mu= 2 \pi$,  since   $\omega_\mu = \{\gamma M, - \gamma M \vec \beta  \} = \{\omega, - \vec k\} $,  \cite{Broglie:1924}.

 For massless particles  (infinite Compton time) these space-time recurrences are potentially infinite (in the  low energy limit).  Very low energy  massless particles ($\lambda^\mu \rightarrow \infty$, for $\mu = 0,1,2,3$) potentially provide the reference non-compact space-time  frames with respect to which massive particle dynamics are described by an \textit{observer}. This allows us to interpret the evolution of massive particles in the whole space-time despite their cyclic space-time dynamics. A massive particle can be located in every non-compact space-time point $x$ defined, for instance, by a low energy massless particle.  In other words particles can propagate in the whole space-time but, point by point, they behaves as embedded in compact space-time  (like bubbles moving on a liquid surface).
 
Each particle field $\Phi(x)$ can be defined on the whole non-compact space-time.  However its evolution in each space-time point $x$ is characterized by intrinsic space-time recurrence  $\lambda^\mu$.  In the particular free particle case  the space-time quantum recurrence $\lambda^\mu$ is \textit{global}   since its four-momentum is also \textit{global} (\ie persistent). In the free case the space-time recurrence does not depend on the particle location point $x$.  When we will describe particle interactions we will promote these space-time recurrences from \textit{global} to \textit{local} quantities.   

% --- so they are not independent each other and the fundamental topology of the cyclic space-time evolution resulting from the FPI  is $\mathbb S^1$.  

%Every cyclic path in the FPI eq.(\ref{path:sum:spacetime}) are degenerate solutions. Since the quantum recurrence $\lambda^\mu$ is fixed by the particle four-momentum $\omega_\mu$, 

Finally we see that the paths in the FPI eq.(\ref{path:sum:spacetime}) are indeed all the \textit{classical} degenerate solutions  linking the final and initial space-time points ${x_i}^\mu $ and ${x_f}^\mu$ --- modulo periods $\lambda^\mu$ --- of the \textit{world-circle} action ($\dot x^\mu (\tau) = d x^\mu/ d \tau$) %\footnote{ Notice that no Lagrange multiplier is necessary in this description, as the constraint $\omega_\mu \omega^\mu = M^2$ is already fixed by the compactification length  of the \textit{world-cycle}.}   
\begin{equation}% \label{Vibrating:String:4D}
\mathcal S_{Compact} = - \frac{1}{T_C }\oint_{(T_C),\tau_i}^{\tau_f} \!\! d \tau \sqrt{\dot x^\mu(\tau)  \dot x_\mu(\tau)}\,. \label{SComp:genericframe}%\\ ~~~&&(\text{ with compactification length $\lambda_C$ and PBCs on $s$})\nn \,.
\end{equation}
Remarkably, the compact space-time description of particle dynamics conciliates QM (here in the Feynman formulation) with the variational principle of \textit{classical} mechanics.   
The Feynman paths are dual to the \textit{classical} vibrational modes of a \textit{world-cycle} string\footnote{Notice that $\mathcal S_{Compact}$ already contains the particle mass as lagrangian constraint, in terms of \textit{world-circle} intrinsic periodicity $T_C = 2 \pi / M$ in $\oint_{(T_C)}$, for the correct relativistic dispersion relation $(\omega_\mu \omega^\mu - M^2) \equiv 0$.}.   

Our result also shows that the target space-time of ECST is the ordinary relativistic four-dimensional space-time. The world-cycle strings describing the elementary particles are therefore characterized by intrinsic space-time periodicities $\lambda^\mu$.  Contrarily to the OST, no ``problematic'' extra-dimensions are needed for the self-consistency of ECST. 

For a free quantum particle of persistent momentum $\vec k$ it is easy to see that the cyclic paths in the FPI eq.(\ref{path:sum:spacetime}) --- related to the  \textit{elementary cycles string} vibrational modes --- interfere constructively if the end points  $x_i^\mu$ and  $x_f^\mu$ of the particle evolution are along the ordinary classical-relativistic trajectory. This is the ordinary point-particle relativistic trajectory in non-compact space-time making stationary the classical-relativistic action $\mathcal  S^{}$ of the FPI, eq.(\ref{Feyn:Path:Int}), proving that  $\mathcal  S^{}$ emerges from the quantum-relativistic action $\mathcal S_{Compact}$ in the effective, classical-relativistic (non-quantum) limit.
On the other hand,  the interference becomes more and more destructive as the sum over cyclic paths is evaluated with one of the end points placed away from the classical trajectory, being $\lambda^\mu$ fixed by $\omega_\mu$. This corresponds to a lower probability to observe the particle away from the classical trajectory,  in agreement with the ordinary probabilistic description of the FPI, \cite{Dolce:2009ce,Dolce:tune,Dolce:AdSCFT,Dolce:cycles,Dolce:2016rfc,Dolce:SuperC,Dolce:EPJP,Dolce:FQXi}.  In the dual stringy description this means that the \textit{world-cycle} string resonates when it is on-shell. 

 In the non-relativistic  limit the sum over paths eq.(\ref{path:sum:spacetime}) reduces to a single Dirac delta over the point-like particle classical path.  This can be easily seen by noticing that the spatial separation between the cyclic paths tends to infinity.  In fact, this limit for massive particles means $|\vec k| \ll M$, that is  $|\vec k | \rightarrow 0$. In the motion direction  the spatial recurrence becomes infinite in this limit, whereas the spatial recurrences are always infinite in the two perpendicular spacial directions, as the momentum components in the perpendicular directions are always zero. Most important, the compactification length along the relativistic time dimension tends to zero in this limit:  $T_C \rightarrow 0$. From the quantum realm characterized by strings vibrating in compact space-time, in the non-relativistic limit we get the classical (galileian,  non-relativistic, non-quantum) point-like particle living in the non-compact cartesian 3D.

Our analysis of the FPI suggests a possible unified scenario of relativistic and quantum dynamics, \cite{Dolce:2009ce,Dolce:tune,Dolce:AdSCFT,Dolce:cycles,Dolce:2016rfc,Dolce:SuperC,Dolce:EPJP,Dolce:FQXi}. 
It seems to realize Einstein's intuition of  his latest years - after Bell's work - when he conjectured that the unification of relativity and QM would be possible by  ``super-imposing''  some sort of BCs to relativistic dynamics \cite{Einstein:1923,Pais:einstein}. Relativity essentially defines the differential structure of space-time without concerning about ``what happens at the boundary of space-time'' whereas it was clear since its early days that QM concerns about  BCs (\eg, particle in a box, Bohr atom, etc).   Summarizing, we have inferred directly from the FPI that the Compton periodicity of elementary particles can be  ``super-imposed''   to the minkowskian space-time geometry, retrieving QM as proven by ECT.  

  From a mathematical point of view the introduction of  space-time BCs fixed by the particle dynamical state, \ie by the particle's four-momentum, together with the ordinary relativistic equations of motion coming from the underlying Minkowskian metric, defines a complete Cauchy problem of relativity itself (relativistic differential structure of space-time dynamics plus intrinsic space-time BCs) solving simultaneously relativistic  and quantum dynamics. The minkowskian metric of ECT preserves the ordinary relativistic differential structure while the controvariant BCs of ECT  quantizes the relativistic dynamics (\eg see the equivalence with the FPI) without breaking relativity (as general principle, the BCs preserve relativistic dynamics as long as they are allowed by the variational principle for that relativistic action, as for the PBCs of scalar relativistic particles investigated here).

  Actually, one of the most interesting consequences of Time Crystals success is that they finally make the assumption PBCs along the relativistic time dimension  a viable physical hypothesis, similar to the ordinary spatial BCs normally used in QM.  PBCs at particle's Compton scales is a natural choice. We already know from the wave-particle duality that elementary particles are ``periodic phenomena'' with the intrinsic space-time periodicities here inferred from the FPI  (BCs at Planck scale however could be relevant for the problem of quantum gravity).  In this way the quantization of relativistic particles is achieved semiclassicaly as a sort of quantization of particles in \textit{elementary space-time boxes} --- or particles in \textit{elementary space-time crystals}, see below, par.(\ref{new:phys:BQM}).

\section{Interactions, QED and Gauge/Gravity Correspondence}\label{interactions}

It is necessary to introduce particle interactions with some detail in order to better understand how causality and the complexity of real physical systems are conciliated within a theory in which elementary particles are intrinsically periodic.  A complete description with extended proofs and formalism is mainly  given  in   \cite{Dolce:tune}. 

Contrarily to the free case, the four-momentum $\omega'_\mu(x)$ of an interacting particle varies \textit{locally} as a function of the space-time point $x$ on which the particle is located, in analogy with the \textit{eikonal} of geometrical optics or modulated signals.  $\omega'_\mu(x)$ is no longer \textit{persistent} but it varies along the particle evolution depending on the type of interaction considered.  On the other hand, the \textit{locally} varying four-momentum $\omega'_\mu(x)$ of the interacting particle fixes \textit{locally} its quantum space-time recurrence in $x$ through the Planck constant. In wave-particle duality the \textit{local} four-momentum $\omega'_\mu(x)$ and the \textit{local} space-time recurrence $\lambda'^\mu(x)$ are indeed two faces of the same coin. 
We must thus promote the \textit{global} space-time recurrence $\lambda^\mu$ of the free case  to a \textit{local} space-time  recurrence $\lambda'^\mu(x)$ in order to describe the \textit{local} four-momentum $\omega'_\mu (x)$ of the interacting particle in $x$. This is achieved by locally deforming the particle space-time compactification lengths, as it can be also inferred directly from the FPI, similarly to par.(\ref{FPI:proof}), see \cite{Dolce:tune}.    In other words the elementary cycle strings associated to particles have locally modulated phases $\omega'_\mu(x)$ during interactions. 

This is a central point to retrieve causality. An observer can establish a ``before'' and a ``after'' the \textit{local} interaction. The effect of interaction is a transition from an initial \textit{local} recurrence to a final \textit{local} one, exactly as in mechanics it denotes a \textit{local} variation of four-momentum.  Now we understand how   the assumption of intrinsic periodicity  in ECT does not mean that the world should be periodic or cyclic as much as the principle of inertia in newtonian mechanics does not implies that everything moves in straight lines. The whole complexity of ordinary physics can be explained in terms of local modulations of periodicities! 

In short, \cite{Dolce:tune}, the full description of interactions in ECT is achieved in terms of space-time geometrodynamics. We want to \textit{locally} deform the elementary space-time cycle compactification lengths in order to obtain local variations of four-momenta. We thus assume a \textit{local} deformation $d x^\mu \rightarrow d x'^\mu(x) = d x^a {e_a}^\mu(x)$  parametrized in terms of the  \textit{vierbein} (\textit{tetrad}) ${e_a}^\mu(x)$. Again,  $x$ is the interaction point on the reference non-compact space-time described above. This, together with the resulting \textit{local} deformation of the compact space-time boundary, yields a \textit{local} modulation of space-time periodicity  $\lambda^\mu \rightarrow \lambda'^\mu(x) = \lambda^a {e_a}^\mu(x)$.  In turn it implies a \textit{local} modulation of phase $\omega_\mu \rightarrow \omega'_\mu(x) =  {e^a}_\mu(x)\omega_a $, that is the  \textit{local} modulation of four-momentum  that we wanted to achieve in order to describe interactions. Thus every interaction can be generally introduced by \textit{locally} deforming the minkowskian space-time (flat) metric associated to the free case $\eta_{\mu \nu}$ to a \textit{local} one: $\eta_{\mu \nu} \rightarrow g'_{\mu \nu}(x)=  {e^a}_\mu(x) {e_\nu}^b(x) \eta_{a b} $, similarly to general relativity. 
The \textit{local} relativistic action $\mathcal S_{Compact}$ describing an interacting particle is therefore defined on a cyclic minkowskian space-time orbifold encoding the local modulations of space-time periodicity of the specific interaction scheme.  
 
 One can easily realize that two types of geometrodynamics are possible \cite{Dolce:tune,Dolce:cycles,Dolce:AdSCFT}. 
 %It implies a geometrodynamical description of the minkowskian cyclic space-time similar: %.  We demonstrate in \cite{} that the local modulations of periodicity associated to interaction are obtained by locally deforming the space-time compactification lengths of ECT: 
The first type, the most familiar one, is characterized by local deformations of the metric tensor leading to a \textit{curved} space-time. Of course this type of elementary cycles local deformations describes gravitational interaction exactly as in ordinary general relativity. For instance, it reproduces the ordinary \textit{clock} rate modulations and \textit{ruler} contractions encoded in a Schwarzschild metric or other general relativity problems.

Remarkably, the second type of deformations reveals a geometrodynamical origin of gauge interactions analogous to that of general relativity \cite{Dolce:tune,Dolce:cycles,Dolce:AdSCFT},  similar to  Weyl's proposal. Due to the compact nature of space-time in ECT it is possible to introduce peculiar interaction schemes (local modulation of space-time phases), that is  particular \text{local} variations of space-time recurrences $\lambda'^\mu(x)$, obtained by locally transforming the metric tensor in such a way that the only effects are local rotations of the space-time boundaries while the metric stays flat (local isomorphism).   
%The local ``rotations'' of the space-time boundary of ECT leaving the metric flat turns out the describe ordinary gauge interactions. 
In spite of the flat metric, in ECT such local rotations of the space-time boundary  corresponds as well to particular \text{local} variations of space-time recurrences $\lambda'^\mu(x)$ and thus to particular kind of interactions that turn out to have exactly the same properties of ordinary gauge interactions (covariant derivatives, gauge invariance, \textit{etc}.), \cite{Dolce:tune}. 

This peculiar type of space-time geometrodynamics associated to gauge interactions have no effect on ordinary fields where space-time has no boundary (they are in fact associated to Killing vector fields). This explains why in ordinary QFT gauge invariance must be necessarily postulated in order to introduce gauge interactions and all the attempts to infer gauge interactions  from geometrodynamical considerations have failed. On the contrary in ECT gauge interactions and gauge invariance can be explicitly deduced from space-time geometrodynamics associated to local rotations of the space-time BCs preserving \textit{locally} the flat minkowskian metric. % They actually imply local modulations of recurrences which can be written in terms of the local variations of four-momenta of gauge interactions (minimal substitution).  
%  As soon as we take into account QM we must consider that  With these elements at hand it is straightforward to conclude that interactions must be introduced in ECT by 
  Particularly simple is  the abelian case (here we only mention bosonic QED). This corresponds to   local $U(1)$ rotations of compact space-time boundary of ECT, such that  ${e^a}_{\mu}(x) = \delta^a_\mu - e {\xi^a}_\mu(x)$ where  ${\xi^a}_\mu(x) \in U(1)$ ---  these are peculiar Killing vector fields associated to  $U(1)$. 
   %described by local variations of space-time periodicity $\lambda^\mu(x)$ such that 
 It results in local modulations of space-time phases  which can be formally expressed in terms of covariant derivatives $D_\mu = \partial_\mu - i e A_\mu(x)$ (``periodicity tuning'') and finally as the ordinary minimal substitution of electromagnetism $\omega_\mu(x) = \omega_\mu - e  A_\mu(x)$, where $e$ is the charge  of the particle and  $A_\mu(x) = {\xi^a}_\mu(x) \omega_a$ is the gauge fields of ordinary electromagnetism. In  \cite{Dolce:tune} we have proven that this description automatically implies all the properties of electromagnetism such as Maxwell equations and gauge invariance;  by following the line of par.(\ref{FPI:proof}), the ordinary FPI of QED is obtained\footnote{Clearly, the condition of intrinsic periodicity (in $x$), which in the global case is $\oint_x {{\omega_n}_\mu} dx^\mu = {{\omega_n}_\mu} \lambda^\mu :=: 2 \pi n$, in the generic interacting case yields the Bohr-Sommerfeld quantization condition $\oint_{x} {{\omega_n}_\mu}(x') dx'^\mu :=:  2 \pi n$, whereas in the particular case of gauge geometrodynamics it will result in the quantization condition of a Dirac string $\oint_x e {A_n}_\mu(x') dx'^\mu :=:  2 \pi n$ and the ordinary scattering matrix of electromagnetism, \cite{Dolce:tune}.}. 
 
%Generalizing the  result of eq.(\ref{path:sum:spacetime}), the FPI of bosonic QED turns out to describe  a sum of cyclic paths with locally modulated  phases. That is, the elementary cycles strings interacting under this peculiar interactions scheme have the local modulations of phases  $\omega_\mu (x)$ of ordinary electromagnetism. 
%It is possible to prove the following identity for the FPI of the abelian gauge interaction (bosonic QED)  \cite{Dolce:tune}:
%\begin{equation*}
%\mathcal Z_{U(1)} =  2 \pi   \sum_{n' \in \mathbb Z} \delta \left(\int_{{x_i}^\mu}^{{x_f}^\mu}\!\!\!(\omega_\mu -  e A_\mu(x)) dx^\mu  + 2 \pi  {n'} \right) \,. 
%\end{equation*}
%% The result leads to a geometrodynamical description of gauge interactions analogous to that of gravitational interaction in general relativity, see \cite{Dolce:tune}. 
% %Gauge invariance is directly inferred from the peculiar geometrodynamics of the compact space-time boundary according to . 
% It must be said that the common geometrodynamical description of gauge and gravitational interactions allowed by ECT shed a new light on the problem of quantum gravity, \cite{Dolce:tune,Dolce:cycles,Dolce:AdSCFT}. 

\section{Predictions of New Physics Beyond Quantum Mechanics}\label{new:phys:BQM}

%It may be surprising that the scale of new physics deduced here from the FPI is at Compton scales (whereas it is common practice to assume new fundamental physics at Planck scale).  But, are we sure to know all the possible physical dynamics up to, say, the LHC energy scale? 

%Physics should have accustomed us with unexpected twists. 
It is definitively true that physicists have explored natural phenomena up to the LHC energy scale. However the experiments were only interested in the \textit{probabilistic amplitudes} associated to the elementary particles. They of course confirm the unquestionable correctness of QM probabilistic predictions of observables, according to Heisenberg, see also \cite{hooft2020fast}, besides testing the Standard Model. In particular the time resolution of LHC or similar experiments is of the order of the picoseconds ($10^{-12}$ sec), ten orders of magnitude below the electron Compton time.  We want here to address the following question. What kind of particle dynamics would we observe if we provide LHC (or more naively an QED experiments) with ultra-accurate timekeepers faster than the Compton scales involved?   

% We have inquired physics at those scales from a statistical point of view,  %But the predictions of QM  are probabilistic.  

Our result suggests something  unedited, while keeping the mathematics of QM perfectly valid for the calculation of probabilistic amplitudes at ordinary time scales. The point is that ultra-accurate timekeepers of resolution better than $T = 2 \pi /\omega $ are necessary to explore possible time dynamics beyond QM characterized by energy scales $\omega$.
The new physics inferred here from the FPI can only be explored by means of ultra-accurate timekeepers able to detect and investigate the ultra-fast cyclic dynamics of elementary particles.

Our arguments suggests that, for instance, new physics beyond ``pure'' QED (\ie limited to electrons interacting electromagnetically) can only be detected by investigating the related quantum phenomena with time accuracy better than %$10^{-21}$ seconds, that is as soon as we will be able to observe electrodynamics with time accuracy better than 
the electron Compton time:  $T_{C_{e}} = 8.09330093 \cdot 10^{-21} sec$.  The electron could reveal its fundamental cyclic string nature if observed with clocks of Compton time accuracy. It is reasonable to hope that this experimental time accuracy will be reached in the next future as the related technologies are already approaching  the atto-second accuracy \cite{Campbell_2017,whurt:2005}.  

We have restricted this prediction to pure QED  experiments. 
%if our experiment involves only electrons interacting electromagnetically it could be sufficient  timekeepers of resolution of the order of the electron Compton time to observe new physics. However, 
If the experiment involves heavier particles of the Standard Model (\eg muons), say of mass $M_q$,  the exploration of new physics beyond QM  would require timekeepers of higher time accuracy $T_{C_q} = 2 \pi / M_q$.  Clearly, the heavier or the more energetic the particle, the faster the new physics cyclic dynamics, as $T = 2 \pi / \omega$.  For instance, the time resolution necessary to explore new physics beyond QM in electroweak dynamics is of the order $T_C^{EW} \sim 10^{-24} sec$ (where we have used the vacuum expectation value $v = 246$ GeV as reference energy). The extreme time resolution necessary to fully explore the  cyclic  dynamics beyond QM in LHC events would be  $T_C^{LHC} \sim 10^{-27} sec$ (where we have assumed the energy of $10$ TeV as reference value).  

For what which concerns neutrinos the present timekeeper resolution would be already sufficient to explore their cyclic dynamics. Indeed we already know that neutrino oscillations are determined by their tiny masses according to the relations described in this paper.  Nevertheless they interact very weakly so it is extremely difficult to investigate the cyclic dynamics beyond QM for the evolution of single neutrinos, as it would be required by ECT. 

As proven in \cite{Dolce:SuperC,Dolce:EPJP}, the dynamics of the charge carriers in Carbon Nanotubes (CN) simulate the  cyclic dynamics  investigated so far at slower time scales than the electron Compton time. The cyclic space-time dynamics characterizing the quantum behavior of CN could be already accessible for present timekeepers.   The compactified dimension of a graphene layer to form a CN plays the role of the world-line compactification  (Compton clock). The charge carriers, otherwise massless in graphene layers (infinite world-line), acquire an effective mass when the CN is formed by dimensional compactification (compact world-line). The resulting effective mass is actually determined by the  compactification length (in this case, the tube circumference) according to the ordinary Compton relation (consider that the speed of light is rescaled in graphene).   The charge carriers at rest along the residual spatial dimension (axial direction) of the 2D CN space-time have actually an effective proper-time periodicity given by the CN diameter. It corresponds to an effective intrinsic proper-time periodicity which actually explains the CNs quantum behaviors and electronic properties in a very straightforward way, in perfect agreement with ECT.  This suggests that CNs could provide an indirect way to study the possible cyclic dynamics beyond QM.   

\section{Emerging Quantum Mechanics from Ultra-Fast Cyclic Dynamics}\label{emerging:QM}

If the time resolution of a quantum experiment is poorer than the Compton time scales of the involved particles, the ultra-fast cyclic dynamics can only be described indirectly in a statistical way. The resulting probabilistic amplitudes are exactly described by the same mathematical rules of ordinary QM \cite{Dolce:2009ce,Dolce:tune,Dolce:AdSCFT,Dolce:cycles,Dolce:2016rfc,Dolce:SuperC,Dolce:EPJP,Dolce:FQXi,Dolce:IntroEC,Dolce:ICHEP2012,Dolce:QTRF5,Dolce:TM2012}, according to the equivalence eq.(\ref{path:sum:spacetime}). Other authors have also pointed out that QM could emerge as probabilistic, low time resolution description of ultra-fast dynamics, see \cite{'tHooft:2001fb,Hooft:2014kka,hooft2021explicit,hooft2020fast,Wetterich_2021}. In particular, an appropriate analogy is that of the outcomes of rolling dice which can only be described probabilistically if observed without ultra-fast timekeepers, see also the analogy with continuous periodic Cellular Automata below.  Roughly speaking elementary particles can be imagined as ultra-fast cyclic ``universes'' interacting each other.

We have rigorously proven in previous papers, \cite{Dolce:2009ce,Dolce:tune,Dolce:AdSCFT,Dolce:cycles,Dolce:2016rfc,Dolce:SuperC,Dolce:EPJP,Dolce:FQXi}, that the low time resolution description of ECT ultra-fast dynamics, at effective level is mathematically and phenomenologically equivalent to ordinary QM in \textit{all} its fundamental aspects. For instance, besides the equivalence with the FPI eq.(\ref{path:sum:spacetime}) proven in this paper, the  equivalence between elementary cyclic dynamics and quantum dynamics has also been proven  for all the fundamental identities of QM, including the canonical formulation of QM  --- the ``Copenaghen axioms'' --- and extended to QFT. All the postulates of QM, including the commutation relations\footnote{In compact dimensions,  integration by parts  implies boundary terms which cancels due to the PBCs yielding exactly the non-commutativities of ordinary QM, in \cite{Dolce:2009ce,Dolce:tune,Dolce:AdSCFT,Dolce:cycles,Dolce:2016rfc,Dolce:SuperC,Dolce:EPJP,Dolce:FQXi} and forthcoming papers.} and Born relation, have been mathematically derived  from the condition of particles' intrinsic periodicity. ECT implies the same mathematical laws of ordinary QM ($ A \Rightarrow B$), the mathematical laws of QM correctly describes all quantum phenomena ($ B \Rightarrow C$), thus ECT can potentially describe quantum phenomena as much as ordinary QM ($ A \Rightarrow C$). 

In ECT there are not hidden variables of any sort, so Bell's or similar no-go theorems cannot be invoked to rule out its equivalence with QM! The ultrafast cyclic dynamics are not driven by hidden variables. They are driven by the ordinary relativistic time, in particular by BCs of the compactified time coordinate determined exclusively by the particle energy and the Planck constant. In other words the ``hidden variable'' would be the relativistic time coordinate, which, of course, is not  a hidden variable. ECT does not obey Bell's inequality as much as ordinary QM. On the other hand the PBCs along the time coordinate is a strong element of non-locality which can be used to interpret coherent states and which conciliates the non-local aspects of QM and the request of locality of relativity  \cite{'tHooft:2001fb,Hooft:2014kka,hooft2021explicit,hooft2020fast,Wetterich_2021}.     

The Feynman paths, written as in eq.(\ref{path:sum:spacetime}) or eq.(\ref{FPI:paths:rest:intro}), are on-shell paths. They are in fact a sum of Dirac deltas. This is indication of ``onticity'' as pointed out by 't Hooft. Actually,  the evolution law of ECT can be alternatively written in terms of 't Hooft ``ontic'' coordinates as $| \tau_i \rangle \rightarrow | \tau_f ~~\text{modulo} ~~ T_C \rangle$, see  eq.(\ref{FPI:paths:rest:intro}), or  $| {x_i}^\mu \rangle \rightarrow | {x_f}^\mu ~~\text{modulo} ~~ \lambda^\mu \rangle $, see eq.(\ref{path:sum:spacetime}),  similarly to continuous periodic Cellular Automata (``particles on a circle'') \cite{'tHooft:2001fb,Hooft:2014kka,hooft2021explicit,hooft2020fast,Wetterich_2021}. This provides an elegant interpretation of the Born relation in terms of elementary cycles\footnote{Similarly to electric currents, \textit{one} particle moving very fast on a circle (circuit), even if neutral, is macroscopically described by a density current and thus by a probability density $|\phi(x)|^2$ \textit{s.t.} $\int dx |\phi(x)|^2 \equiv 1$. } and, according to 't Hooft arguments, it suggests that the new physics beyond QM predicted by ECT is potentially \textit{deterministic}, \cite{Dolce:2009ce,Dolce:tune,Dolce:AdSCFT,Dolce:cycles,Dolce:2016rfc,Dolce:SuperC,Dolce:EPJP,Dolce:FQXi}.  

In ECT the space-time can be equivalently imagined as forming Compton elementary space-time cells or boxes, due to the condition of periodicity in space and time. Therefore we find a strong relationship between ECT and Quantum Time Crystals \cite{Wilczek:2012jt}. These elementary space-time cells can be imagined to form \textit{elementary space-time crystals}. In ECT the BCs along the time direction determine the quantum behaviors of the elementary particles rather than of the atomic crystals. We can summarize by saying that in ECT the full quantization of elementary particles is achieved as relativistic generalization of the ``semi-classical''  quantization (Bohr-Sommerfeld), \ie as sort of particles in space-time ``boxes''. 

\section{Form the Intrinsic Time Periodicity of  ``Pure'' Quantum Systems to the Entropic Arrow of Time of Thermodynamical Ensembles}\label{arrow:time}

We discuss how ordinary thermodynamics and the related arrow of time emerge from a description of physics in which elementary isolated particles are assumed to be intrinsically periodic in time.  Notice that, in any case, the particles' cyclic behavior is essentially that prescribed by ordinary undulatory mechanics, although with the fundamental difference that in ECT these cyclic dynamics originate directly from the fabric of space-time. Elementary particles are the elementary clocks of nature, so we can say that they are characterized by ``complete coherence'' (minimal entropy) if isolated by the rest of the universe. 

 We recall the meaning of the temperature $\mathcal T$ in thermodynamical systems by introducing the thermal time  $\beta = 1 / k_B \mathcal T $, where $k_B$ is the Boltzmann constant, which is also known as  Euclidean time (intrinsic) periodicity. 
Again, the central point to bear in mind is that the intrinsic Minkowskian time periodicity $T = 2\pi / \omega$ described so far for elementary particles and their energy $\omega$ are two faces of the same coin, exactly as $\beta$ and $\mathcal T$. As already noticed in par.(\ref{interactions}), interactions involving for instance an exchange of energy $\Delta \omega = 2 \pi / \Delta T$ necessarily implies a local modulation of periodicity $\Delta T = 2 \pi /  \Delta \omega$. In ECT these \textit{local}  modulations result from space-time geometrodynamics.
 
A thermodynamical system at finite temperature is an ensemble with stochastic collisions among particles. The thermal noise therefore  randomly breaks the ``complete coherence'' of the elementary particles with sudden variations of particles' periodic regimes. Since this occurs with a mean time rate of the order of $\beta$ we can understand that the Euclidean  and the Minkowskian time periodicities are in competition.  No surprise that quantum effects are typical of very low temperatures  and  high temperature tends to destroy quantum coherence.  In other words the intrinsic Minkowskian periodicity described so far referees to pure quantum systems, \eg isolated particles at zero temperature, which are indeed perfect coherent states. It represents the typical autocorrelation time of the pure quantum phenomena, in opposition to the thermal time $\beta$ representing the rate at which the quantum correlation is destroyed. 

  If the rate of the thermal noise collisions is sufficiently low, \ie the Euclidean periodicity $\beta$ is larger than the  Minkowskian periodicity, $T \ll \beta$, than the system can auto-correlate and give rise to quantum effects. In the opposite limit, $\beta \ll T$, the thermal stochastic collisions are so fast that they break the quantum recurrence given by the Minkowskian time  periodicity before  autocorrelation phenomena occur. The transition between quantum and thermodynamics (critical temperature) is therefore parametrized by the ratio $\beta / T$.  The effect of the thermal noise is indeed, at macroscopic level, a dumping $e^{- 2 \pi \beta / T} = e^{- \omega/ k_B \mathcal T}$ of the intrinsic cyclic behavior of pure quantum systems. This is  the Boltzmann factor typically appearing in front of the quantum phasor $e^{- i \omega t}$ in finite temperature systems. 

Our results shed a new light on QFT at finite temperature. From the Matsubara theory  we know that the prescription to quantize thermodynamical systems is to perform a Wick's rotation $t \rightarrow -it$ and then to impose  intrinsic periodicity  $\beta$ as constraint to the resulting Euclidean time. This method is often considered as  a ``mathematical trick''. % ECT allows a novel physical explanation to this quantization prescription.  Our analysis of the FPI simply says that Minkowskian time intrinsic periodicity is the origin of quantum phenomena: the quantization condition is the constraint of periodicity.
The opposing roles of quantum mechanics and thermodynamics are well represented by the Wick rotation which transforms the perfect recurrence of a pure quantum system represented by the phasor  $e^{- i \omega t}$  to a dumping of the minkowskian periodicity $e^{- 2 \pi \beta / T} = e^{- \omega/ k_B \mathcal T}$ associated to a finite temperature and resulting in the Euclidean time periodicity. Our analysis of the FPI simply says that Minkowskian time intrinsic periodicity imposed as constriant is the origin of quantum phenomena.  The constraint of Minkwskian time periodicity is a quantization condition for zero temperature systems as much as the constraint of Euclidean time periodicity is a quantization condition for finite temperature systems.  Actually, QFT theory at finite temperature was one of the original motivations for ECT and represents an indirect support to the idea of intrinsic periodicity. As we will mention below this argument has been successfully applied to consistently describe condensed matter quantum phenomena in terms of ECT, such as superconductivity and graphene physics,  see also \cite{Dolce:IntroEC,Dolce:2016rfc}.

Now we speculate about the emergence of the entropic arrow of time from the condition of intrinsic Minkowskian periodicity.  Pure QM systems are ideal systems characterized by unperturbed time periodicity as it can be easily observed in so many quantum phenomena (from Bohr atom to quantum time crystals). This ideal (zero temperature) realm of pure QM   ---  and the possible relativistic physics beyond it described in this paper --- are the fundamental systems of nature. As soon as QM is concerned we must replace the Newton's law of inertia  with the principle of intrinsic periodicity of isolated particles.   The intrinsic periodicity of a zero temperature elementary particle plays the role of straight line motion of the ideal case of free particles in Newton's laws. In our everyday life we experience an chaotic world due to interactions among  the enormous number of elementary particles constituting ordinary thermodynamical systems. The arrow of time emerges for entropic reasons from the ideal quantum realm characterized by intrinsic periodicity as it emerges from the Newton's law as soon as the thermal noise is considered. In both cases the reason is that we must give up with a detailed microscopic description of the thermal collisions among particles and adopt a macroscopic statistical description (kinetic theory of gases). Notice that QM emerges from intrinsic Minkowskian periodicity with a very similar mechanics, except that the necessity of a statistical description is given by the ultra-fast cyclic dynamics intrinsic in each single particle rather than by the enormous number of particles involved.  

At fundamental quantum level the microscopic minkowskian time seems to have very different properties with respect to the macroscopic entropic time. Every elementary particle, if isolated from the rest of the universe,  at zero temperature, is a pure auto-correlated system whose behavior is intrinsically periodic. Time is a periodic minkowskian coordinate from the point of view of an isolated particle.  It is undeniable that pure quantum systems are characterized by precise recurrences in time.  In ordinary physical systems this cyclic aspect of quantum time is destroyed by the thermal noise. Our macroscopic experience of the time flow is therefore dominated by thermal aspects --- even though, it must be said, a cyclic character of time survives in many aspects of nature, from astronomical observations to biology.  The flow of time has to be considered  a macroscopic effect originating from the mathematics of stochastic systems --- as for classical thermodynamics. Even though the elementary particles composing an ensemble are periodic phenomena if isolated from the rest of the universe, the thermal noise implies that at macroscopical level they evolve, for purely statistical reasons and due to stochastic interactions, towards the most probable state when they form a thermodynamical system. Such a ensemble has finite entropy and its evolves according to the second principle of thermodynamics. In ECT every elementary particle is an elementary clock which can be \textit{arbitrarily} imagined to rotate in clockwise direction, fixing in this way the \textit{positive} direction of the microscopic Minkowskian time. If we now imagine to invert all the elementary clock rotations to anti-clockwise, \ie to \textit{invert} the Minkowskian time direction, according to our arguments in which interactions are local modulations of periodicities, we would obtain the vary same macroscopic picture with the same stochastic evolution and the same entropic arrow of time for the system as before\footnote{Since every particle is a clock, the external minkowskian time coordinate can be dropped in ECT and the inversion of a single elementary clock rotation corresponds to transform the particle to anti-particle, similar to Feynman's interpretation. A universe made of antiparticle --- anti-clockwise clocks, that is particle traveling backward with respect to our minkowskian time  --- would experience the same entropic time arrow of our universe.}. So, it seems reasonable to admit that the entropic arrow of time, \ie  the macroscopic flow of time, emerges for stochastic reasons from the intrinsic periodicity of elementary particles, \ie from the microscopic minkowskian time which has a intrinsically cyclic nature, see also \cite{Dolce:FQXi,Dolce:TM2012}. With this argument we have accounted that the principle of intrinsic periodicity is essentially reconcilable with the law of thermodynamics and the entropic time arrow.

\section{Remarks}

%Finally we mention further interesting aspects of the picture of new physics provided by ECT.
ECT is a formulation of elementary particles physics, simple (``but not simpler'' [A. Einstein]), effective, and \textit{refutable} 
in addition to falsifiable. %\footnote{Beside the analogies with Time Crystals and the others, the apparently critical hypothesis such as compact relativistic time at Compton scales can be proven to be either absolutely correct or completely inconsistent with ordinary physics, but notice that it has passed quite a few of peer-reviews \cite{Dolce:2009ce,Dolce:tune,Dolce:AdSCFT,Dolce:cycles,Dolce:2016rfc,Dolce:SuperC,Dolce:EPJP,Dolce:FQXi,Dolce:IntroEC,Dolce:ICHEP2012,Dolce:QTRF5,Dolce:2010ij,Dolce:TM2012} where the many demonstrations of its full consistence are scientifically certified. All this is sign, at the very least, that the matter is not so obvious. As Feynman used to say: ``\textit{Doubt is clearly a value in science. It is important to doubt and the doubt is not a fearful thing, but a thing of great value}''. This paper want to promote  a serious discussion on the scientific merit  of the compact space-time hypothesis, as really deserved.} 
In fact, it does not involve extra parameters (often characterizing fine-tuning plays in modern physics) or hidden variables, being based exclusively on the ordinary elements of relativity (space-time coordinates) and the use of the Planck constant.  
%We summarize essential facts about ECT, \cite{Dolce:2009ce,Dolce:tune,Dolce:AdSCFT,Dolce:cycles,Dolce:2016rfc,Dolce:SuperC,Dolce:EPJP,Dolce:FQXi,Dolce:IntroEC,Dolce:ICHEP2012,Dolce:QTRF5,Dolce:2010ij,Dolce:TM2012}. 
More that ten years of researches,  \cite{Dolce:2009ce,Dolce:tune,Dolce:AdSCFT,Dolce:cycles,Dolce:2016rfc,Dolce:SuperC,Dolce:EPJP,Dolce:FQXi,Dolce:IntroEC,Dolce:ICHEP2012,Dolce:QTRF5,Dolce:TM2012}, supported by solid and crosschecked mathematical demonstrations and nearly 20 published papers on the topic, mostly peer-reviewed, suggest that a cyclic (or more in general compact) formulation of space-time at Compton scales is definitively a viable hypothesis (as indirectly shown by Time Crystals). Despite the controversial hypothesis of intrinsic time periodicity, in the last analysis ECT turns out to be absolutely consistent with all the theoretical foundations and the essential phenomenology of both quantum physics and  relativity (special and general), \cite{Dolce:2009ce,Dolce:tune,Dolce:AdSCFT,Dolce:cycles,Dolce:2016rfc,Dolce:SuperC,Dolce:EPJP,Dolce:FQXi}. It also reveals a fundamental  relationship between Feynman's and de Broglie's based interpretation of QM; between the ``new'' and the ``old'' formulations of QM.

In addition to the remarks already given in this paper,  we mention that ECT is supported by many other foundational ideas of theoretical physics at the origin of the Kaluza-Klein theory, the Kaluza miracle, Holography, the $AdS-CFT$ correspondence, Loop Quantum Gravity, just to mention a few.  A remarkable result,  mainly proven in  \cite{Dolce:AdSCFT}, is that in ECT the ``quantum to classical correspondence'' \cite{Witten:1998qj} at the base of the AdS/CFT correspondence has been proven to be an exact mathematical identity rather than a conjecture. Consider for instance the equivalence presented in this paper between quantum particle dynamics described by the FPI and  classical cyclic string dynamics. Moreover,  ECT is dual to an extra-dimensional theory where the compact world-line plays the role of a ``virtual extra-dimension'' with interesting insights into the Kaluza's miracle. The common geometrodynamical description of gravitational and gauge interactions mentioned above, see  the detailed proofs in \cite{Dolce:tune}, can be actually regarded as  a Gauge/Gravity correspondence \cite{Dolce:AdSCFT}. 
ECT is manifestly holographic \cite{Dolce:tune,Dolce:cycles,Dolce:AdSCFT}: interactions, and the consequent local modulations of space-time recurrences, can be equivalently described as deformations of the space-time compactification lengths. Particle dynamics are encoded into the shape of the particles' space-time boundary. 

The success of ECT is not limited to particle theoretical physics. It has also been confirmed, for instance, in condensed matter where it has been used to produce fundamental phenomenological results.  
ECT has been successfully applied to infer \textit{all} the fundamental aspects of superconductivity and graphene physics (mentioned above) directly from the physical principle of intrinsic periodicity, rather than from empirical models (\eg, as for the BCS model), \cite{Dolce:SuperC,Dolce:EPJP}. In particular the Josephson effect, which is one of the most striking examples of Time Crystals, has a straightforward explanation in ECT. It is in fact a direct consequence of periodic BCs in time applied to Maxwell's laws.  ECT promotes intrinsic time periodicity to foundational principle of physics and QM emerges from the resulting constrained (``super-imposed'') relativistic dynamics.

 \section{Conclusions}
 
The FPI can be consistently reformulated in terms of elementary space-time cyclic dynamics,  classical in the essence. Our proof suggests a reconsideration of the fundamental nature of relativistic space-time  for possible physics beyond QM, in which the idea of Time Crystals seems to be extended to elementary particles.  

New physics will be discovered by investigating quantum phenomena with time precision better that $10^{-21} s$. With sufficiently accurate ultra-fast timekeepers elementary particles will reveal their novel stringy nature. Such a novel  Elementary Cycle String Theory  is characterized by  one-dimensional world-lines compactified at Compton lengths named particle  \textit{world-cycles},  rather than by the two-dimensional \textit{world-sheet} of Ordinary String Theory.  It implies an intrinsically cyclic nature of relativistic space-time at the base of QM.  This  is possible because elementary particles are the elementary clocks of nature.  

The  scenario is  also confirmed by many others crosschecked mathematical and phenomenological equivalencies with ordinary quantum physics already peer-reviewed and published in \cite{Dolce:2009ce,Dolce:tune,Dolce:AdSCFT,Dolce:cycles,Dolce:2016rfc,Dolce:SuperC,Dolce:EPJP,Dolce:FQXi}, see also \cite{Dolce:IntroEC,Dolce:ICHEP2012,Dolce:QTRF5,Dolce:TM2012}. We have also discussed the consistence between the intrinsically periodicity of the minkowskian time coordinate of zero temperature free particles or, more in general, perfect coherent states, and the entropic arrow of time of finite temperature thermodynamical ensemble of particles, noticing the deep analogy played as quantization condition by the constraint of intrinsic minkowskian and euclidean periodicity, respectively. 
 
 With all these elements at hand we conclude that the hypothesis of intrinsically cyclic (compact) space-time is conceptually legitimated, mathematically correct and phenomenologically predictive. The result is a  unified formulation of quantum and relativistic physics \cite{Dolce:2009ce,Dolce:cycles,Dolce:2016rfc,Hestenes:zbw:1990}  characterized by no fine-tunable parameters or hidden-variables of any sort and potentially deterministic \cite{'tHooft:2001fb,Hooft:2014kka,hooft2021explicit,hooft2020fast,Wetterich_2021},
 
QM emerge from the constraint of relativistic intrinsic periodicity of particle dynamics as statistical, low time resolution, effective description, in analogy with the statistical (probabilistic) description of  the outcomes of rolling dice observed without slow motion camera. 
``Does God play dice''?  Assuming that ``God'' has potentially infinite resolution in time (infinitely accurate timekeeper), ``He'' would always be able to directly observe the ultra-fast cyclic dynamics beyond QM and, similar to dice observed with a slow motion camera, to predict deterministically the quantum outcomes --- without the fun.  The solution to Einstein's quantum dilemma seems to be that ``God'' has no fun playing quantum dice.  
\newline

\noindent $\mathbf{Acknowledgments}$ I would like to thanks Prof. Ignazio Licata for the useful discussions and comments. 

\providecommand{\href}[2]{#2}\begingroup\raggedright\endgroup

\end{document}